\documentstyle[epsfig,preprint,epsf,aps,eqsecnum]{revtex}
\tightenlines
\newcommand{\dslash}{\partial\hspace{-.09in}/}

\newcommand{\lsim}{\mathrel{\lower4pt\hbox{$\sim$}}
\hskip-12.5pt\raise1.6pt\hbox{$<$}\;}
\newcommand{\gsim}{\mathrel{\lower4pt\hbox{$\sim$}}
\hskip-12.5pt\raise1.6pt\hbox{$>$}\;}

\begin{document}
\title{Neutrino ground state 
in a dense star}

\author{Ken Kiers\footnote{Email: kiers@bnl.gov\vspace{-.15in}}}
\address{High Energy Theory, Department of Physics\\ 
Brookhaven National Laboratory,
Upton, NY 11973-5000, USA}
\author{Michel H.G. Tytgat\footnote{Email: 
mtytgat@muon.phy.bnl.gov\vspace{-.15in}}}
\address{High Energy Theory, Department of Physics\\ 
Brookhaven National Laboratory,
Upton, NY 11973-5000, USA\\
and\\
Service de Physique Th\'eorique\\
Universit\'e Libre de Bruxelles, CP225\\
Bd du Triomphe, 1050 Bruxelles, Belgium}
\maketitle

\begin{abstract}

It has recently been argued that long range forces due to the
exchange of massless neutrinos give rise to a very large self-energy 
in a dense, finite-ranged, weakly charged medium. Such an effect,
if real, would destabilize a neutron star. 
To address this issue we have studied the related problem of
a massless neutrino 
field in the presence of an external, static 
electroweak potential of finite range. 
To be precise, we have computed to one loop the  exact 
 vacuum  energy for the case of 
a spherical square well potential of depth $\alpha$ and radius $R$.
For small wells, the vacuum energy is reliably
determined by a perturbative expansion in the external potential. 
For large  wells, however, 
the perturbative expansion breaks down. A manifestation of this breakdown
is that the vacuum
carries a non-zero neutrino charge. The 
 energy and neutrino charge of the ground state are,
to a good approximation for large wells, 
those of a  neutrino condensate
with chemical potential~$\mu=\alpha$.  Our 
results demonstrate explicitly 
that long-range forces due to the exchange of massless neutrinos
do not threaten the stability of neutron stars.
\end{abstract}

\newpage

\section{Introduction}
\label{sec:intro}

If neutrinos are massless, the exchange of neutrino-antineutrino
pairs gives rise 
to  long range interactions between weakly charged particles, such as
neutrons~\cite{feinberg,hsu}. 
On dimensional grounds, the
potential between two neutrons separated by a distance $r$
 must be  of the form
$$
V \sim {G_F^2 \over r^5} 
$$
(where $G_F$ is the Fermi constant), and thus falls off rapidly at large
distances. 
As was first remarked by Feynman~\cite{dick,hartle},
an interesting effect  may arise if 
there is a finite density of  matter 
near the two neutrons. A 4-body process, for instance, 
can give rise to a potential that behaves like
\begin{equation}
\label{grav}
V \sim {G_F^4 N^2\over R^8}\,{1\over r} ,
\end{equation}
if $R \gg r$, where
$N$ is the total number of particles in the cloud of matter 
and $R$ is the distance between the matter and the two test neutrons.
Because of the large  $N^2$ enhancement factor
in~(\ref{grav}), the effective coupling
$G_F^4 N^2/R^8$ can be of order unity.  In this case,
far-away matter would  give a large
contribution to the two-body potential. 
Since this effect would be  even more important for higher-order 
processes, one would then have to take into account
diagrams of arbitrarily high order.
 
More recently, Fischbach~\cite{fb} has studied 
the contribution to the energy of a neutron star due to the 
exchange of massless neutrinos.
Large combinatoric enhancements, 
similar to those envisioned by Feynman,
led him to conclude that a neutron star could not
be gravitationally bound unless neutrinos
are massive, $m_\nu \gsim 0.4$~eV.\footnote{
Smirnov and Vissani~\cite{smirnov1} 
have argued that the long range forces can also be screened  by
the neutrino condensate that resides in a star~\cite{loeb}. 
However, unlike screening at finite temperature, screening at finite density
is  not perfect. Fischbach, {\em et al}~\cite{fb2} have argued, in response,
that in a finite system a residual effect could still blow up
a star (or make it collapse into a black hole).}

Abada {\em et al}~\cite{abada} 
have  contended  that Fischbach's
calculation was incorrect.
They have concluded that the neutrino contribution to the
energy of a neutron star is essentially negligible.
While the arguments given by Abada {\em et al} were compelling, 
the calculations upon which they were based
did not take into account the finite size of
the star.
As we will demonstrate in this paper, the conclusions 
reached in~\cite{abada} were nevertheless essentially correct.

We believe that we have 
completed the proof of the assertions made in~\cite{abada}.
We have studied the ground state properties of a 
massless neutrino field propagating in
a  finite-range electroweak potential, 
like the one induced
by a finite density of neutrons.  To be precise, we
have considered a spherical square well electroweak potential 
of depth $\alpha$ and radius $R$, and have
computed to one loop the exact energy and   
neutrino number of the ground state.
By exact we mean that our results are 
non-perturbative in the external potential.
In order to carry out this program, we 
have used an approach due initially to 
Schwinger~\cite{julian}.\footnote{Similar calculations 
are found  in non-perturbative computations of  
 quantum corrections to solitons~\cite{neveu,jaffe,bordag} and in
studies of 
QED in strong 
fields~\cite{raf}.}
For technical reasons, the potentials which we 
consider are of relatively 
modest strength and range compared to those of a realistic 
neutron star.  As we shall argue, our calculation
is  nevertheless sufficient to 
demonstrate that the neutrino ground
state energy in a neutron star 
does not vary wildly as the radius or
density of the star is increased,
but is in fact very well-behaved.

The paper is organized as follows.  In Section~\ref{sec:rev}
we introduce the reader to Fischbach's results and
review  the critique advanced by Abada, {\em et al}.
We then make the case for our approach to the question.
We show that we have a well-defined
problem to which we can give a definitive solution.

Section~\ref{sec:scat}
 is essentially independent of the specific application
to the neutron star  problem and, we believe,
is of interest in its own right.
We first solve the Weyl equation in the presence
of a spherical square well potential. Because the
fermions are massless, there are no bound states 
 but only scattering
solutions. From the expression for the 
phase shifts, we compute the properly renormalized
energy of the neutrino ground state in the presence of the potential.
For small wells,
 $\alpha R \ll \pi$, we show
that the energy of the ground state is  perturbative in the potential. 
As the size of the potential increases, the perturbative expansion
 breaks down. A symptom of this breakdown is that for $\alpha R \geq \pi$
 the  ground state
carries a  non-zero neutrino charge. 
We show that, to 
a very good approximation for large wells, the  ground 
state energy and neutrino
number are those of  a neutrino 
condensate with chemical potential $\mu=\alpha$, 
a result  anticipated by Abada {\em et al}. 
That this result also holds  for a neutron star follows trivially.  

We emphasize that 
we have studied precisely the same problem as Fischbach did.
Our approach, however, is quite different. 
We conclude that long range forces
due to the exchange of massless neutrinos do not destablilize a star.
There is indeed a shift in vacuum energy, but -- maybe  unfortunately -- it
is too tiny to be of any  consequence 
regarding the fate of a neutron star.

Related lines of argument can be found in the recent literature. 
In~\cite{recone}, a $1+1$ dimensional star is studied as a toy model.
In this case, the vacuum energy can be calculated exactly and 
happens to vanish.
It is also shown that the presence of a boundary implies the
existence of a neutrino condensate.  This point is also emphasized
in~\cite{rectwo}.  Both papers stress that
a complete $3+1$ dimensional calculation for a system with 
a boundary is necessary in order to settle the issue definitively.

\section{The neutrino ground state in a neutron star}
\label{sec:rev}

\subsection{Preliminaries}
Before embarking on our main calculation,
we introduce some useful concepts and formulas. 
The stated goal in Ref.~\cite{fb} is
to calculate the shift in vacuum energy due to neutrino
exchange in a neutron star:
\begin{equation}
\label{shift}
W = \langle \hat 0 \vert  H \vert \hat 0 \rangle - \langle 
 0 \vert  H_0 \vert  0 \rangle .
\end{equation}
Here $H$ denotes the Hamiltonian in  the medium and 
$\vert \hat 0 \rangle$ refers to the true
ground state of the system. 
The second term  is the usual matter-free
vacuum energy. 
For fermions, Eq.~(\ref{shift}) amounts to 
\begin{equation}
\label{discsum}
W = - {1\over 2} \left \{ \sum_{i > 0}
 (E_i - E_i^0) -
\sum_{i < 0} (E_i - E_i^0)\right \} ,
\end{equation}
where the indices $i$ refer  to
the positive and negative energy levels of the neutrinos.
A first step toward the actual  computation of~(\ref{shift}) 
is the
following expression due to Schwinger~\cite{julian},
\begin{eqnarray}
\label{sch}
W &=& - \int d^3 \! x \,\, {\partial\over \partial
t}\,\mbox{\rm tr}  
\left [S_F(x, x^\prime) - 
S_F^0(x, x^\prime) \right]_{x^\prime \rightarrow x}
\end{eqnarray}
where $S_F(x,x^\prime)$  
is the $2\times 2$ Feynman  propagator for the massless left-handed neutrino
field in the presence of the finite neutron density
and $S_F^0(x,x^\prime)$ is that of the free field.
In~(\ref{sch}), the limit $x^\prime \rightarrow x$ 
 is  taken symmetrically and  averaged. 

Let us now drop any explicit reference to the
neutrons of the star and consider instead the 
following effective Lagrangian for the neutrino field
\begin{equation}
\label{efflag}
{\cal L}_{\rm eff} = 
  \overline{\psi}_{L} \left[i\dslash 
        + \alpha \gamma^0\right]\,\psi_{L}
\end{equation}
where $\psi_L = L \psi$, $L = (1 - \gamma_5)/2$, and where
$$
\alpha(\vec{x}) = G_F\rho_n(\vec{x})/ \sqrt{2}\, \sim 20\,\,{\rm eV}
$$
is the  electroweak  potential 
induced by the finite
neutron density ($\rho_n \approx 0.4\,\, \mbox{\rm fm}^{-3}$ 
in a typical neutron star). 
The static potential $\alpha > 0 $  is attractive for neutrinos and 
repulsive for antineutrinos. 
The effective Lagrangian of~(\ref{efflag})
has been  derived in various ways from the underlying electroweak 
theory~\cite{mann,ken1,ken2} and applied to studies of 
the Mikheyev-Smirnov-Wolfenstein (MSW) effect~\cite{msw}.
It  provides us with  a most convenient tool to study
  the coherent behavior
of low-energy neutrinos  in a neutron star, on scales large 
compared to the size of a neutron
$r_n\sim1\,\mbox{\rm fm}\approx(200\; \mbox{\rm MeV})^{-1} $.\footnote{
The fact that
the effective theory, as in~(\ref{efflag}), is  anomalous 
should not worry us. The underlying theory (the Standard Model)
is perfectly well defined. Besides, for a static
potential in $3+1$ dimensions, there are no anomalous effects
because $\vec E \cdot \vec B = 0 $.}

Since the potential is static, it is convenient
to introduce
\begin{equation}
S_{F}^{(0)}(x, x^\prime) = {1\over 2 \pi}\int_C d\omega \, e^{- i 
\omega (t - t^\prime)} 
\, G_{(0)}(\vec x, \vec x^\prime;\omega)
\end{equation}
with $C$ representing the usual
 Feynman contour. 
The resolvents $G(\vec x, \vec x^\prime; \omega)$ and 
$G_0(\vec x, \vec x^\prime; \omega)$  satisfy
\begin{eqnarray}
(\omega -H_{(0)}) \, G_{(0)}(\vec{x},  \vec x^\prime;\omega) &=& - 
	\delta^3(\vec x - \vec x^\prime)
\end{eqnarray}
with the Hamiltonian  from~(\ref{efflag})
\begin{equation}
\label{h}
H = - \alpha + i \vec\sigma\cdot\vec \nabla_{x};\,\,\,\,\,\,H_0 = i 
\vec\sigma\cdot
\vec\nabla_x.
\end{equation}
The Schwinger formula,~(\ref{sch}), becomes 
\begin{eqnarray}
W & = & {i\over 2 \pi} \int_C d\omega\,
 \,\omega \, \mbox{\rm Tr}_{\bf x}\left [G(\omega) - G_{0}(\omega)\right ]
	\nonumber \\
 & = & -  {i\over 2 \pi} \int_C d\omega\,  \omega\,\mbox{\rm Tr}_{\bf x}
\left [{1\over \omega -H} - {1  \over \omega  - H_0} \right]\label{sch2}
\end{eqnarray}
where the trace is 
over spinorial and configuration 
space indices.\footnote{The rule for closing the Feynman contour $C$
in the complex $\omega$ plane in Eq.~(\ref{sch2}) is to take 
the average of the integral around the 
positive and negative energy cuts, so as to 
recover~(\ref{discsum}).}
Integrating~(\ref{sch2}) by parts and expanding  the logarithm in
powers
of the
external potential finally gives
\begin{eqnarray}
\label{trlogd}
W &=&  - {1\over 2 \pi i} \int_C d\omega\,  \mbox{\rm Tr}_{\bf x}
\log \left [{\omega -H \over \omega  - H_0}\right ]\\
&& \nonumber\\
\label{exp2}
&= &   {1\over 2 \pi i} \sum_{k = 1}^\infty \, {1 \over k}\,
 \int_C d\omega\, 
 \mbox{\rm Tr}_{\bf x} \left  [{\alpha  G_0(\omega)}\right ]^k
\end{eqnarray}

Even with the free particle vacuum energy  subtracted,
 Eq.~(\ref{exp2}) is still  a formal, ultraviolet-divergent 
quantity. The culprit is the first 
term,\footnote{Only the $k-even$ terms contribute 
for a static external potential. This is most easily seen in
 the Dirac representation of~(\ref{efflag}) because 
then the coupling to the potential
involves the $\gamma_5$ matrix and the trace over an odd
number of vertices is proportional to the Levi-Civita tensor.
For a static
potential, there  
 are not enough indices to be contracted with the Levi-Civita
tensor and the odd terms vanish identically.} $k=2$,
which is  the familiar  vacuum polarization
diagram.
This diagram is superficially  quadratically divergent
and has to be renormalized. However, because of 
gauge invariance, the actual ultraviolet divergence is milder. 
We will discuss this point in more detail in 
Section~\ref{sec:scat}. Let us simply mention
here that
$$
\vert W^{(2)}_{\rm ren} \vert < R^3,
$$
strictly. As $R$ increases, the $k=2$ term
 gives a vanishing contribution to the energy per unit
volume of the star.\footnote{This statement is correct for a smooth
effective  potential, $\alpha$. 
In a realistic star, the coarse grained
 structure at the  neutron scale leads to a vacuum polarization contribution 
which scales like $R^3$.  See also the ``Note added in proof.''}

The remaining contributions, $k \geq 4$,  are 
 ultraviolet convergent because the interaction term
in (\ref{efflag}) is renormalizable. 
On the other hand, because the neutrinos are massless,
these terms have a strong infrared dependence
\begin{equation}
\label{correct}
W^{(k)} \sim  {1\over k} \, {1\over R}\, (\alpha R)^k .
\end{equation}
For a neutron star, $\alpha \sim 20$ eV and $R \sim 10$ km,
so that $\alpha R \sim 10^{12}$. 
With such an expansion parameter, 
the  perturbative expansion  of~(\ref{shift}) 
is doomed to diverge  and Eq.~(\ref{exp2}) must be resummed.

For  comparison, the estimate of  the vacuum 
energy~(\ref{shift}) given in~\cite{fb} is a large but finite sum,   
\begin{equation}
\label{fish}
W = \sum_{k = 2}^N W^{(k)} ,
\end{equation}
where $N \sim 10^{57}$ is the number of neutrons in the star.
Each term is taken to be given by the average $k$-body potential energy 
multiplied by the number of ways of choosing $k$ neutrons from
the grand total of $N$.  For $k$ large, but much less than $N$,
\begin{equation}
\label{wrong}
W^{(k)} \sim {G_F^k\over R^{2 k +1}} 
\times 
\left(\begin{array}{c}
N\\
k\\
\end{array}\right)
\sim {1\over k !} \, {1\over R} \, \left 
({G_F N\over R^2}\right )^k \sim {1\over k!}\,{1\over R} \, (\alpha R)^k .
\end{equation}
Since $\alpha R \sim 10^{12}$, it is not surprising that 
after only a few terms the sum in Eq.~(\ref{fish}) leads
to an estimated energy which exceeds the  mass of the neutron 
star.  For future reference, let us quote the following 
 estimate obtained in Ref.~\cite{fb2}. 
For fixed neutron  density, corresponding to  $\alpha \sim 20$~eV,
a subsystem of  radius $R \sim 2\cdot10^{-5}$~cm
 contains as much energy as the 
total mass of a neutron star with a radius of $10$~km.
Note also that 
$\alpha \, R = 20\, {\rm eV} \times 2\cdot 10^{-5}\, {\rm cm} \approx 20$.

 Although similar, the 
estimates~(\ref{correct}) and~(\ref{wrong}) differ 
crucially in the details. As emphasized by Abada, 
{\em et al}~\cite{abada}, for fixed $k$, there are
${\cal O}[(k-1)!]$  terms missing in~(\ref{wrong}). 
These  are associated with multiple rescattering over
the same neutrons. These processes also turn 
the finite sum~(\ref{fish}) into an infinite
series, as in~(\ref{exp2}). For large $\alpha R$, the series is
diverging and must be resummed.

\subsection{The scattering problem}

Consider the  expression given in~(\ref{discsum}) for the vacuum energy. 
We can count the energy levels of the
Hamiltonian~(\ref{h}) by putting the system in a large box~\cite{julian}.
Let us  specify each eigenstate  by its
 energy eigenvalue $E$ and by
a set of quantum numbers $\kappa$. 
For simplicity, consider a spherically symmetric
potential in a large spherical box of radius $R_{\rm box}$. 
(The parameter $\kappa$
is then related to the total angular momentum.)
Because the neutrinos are
massless, the spectrum of~(\ref{h}) has  no bound states 
but only scattering solutions,
\begin{equation}
\psi_{L\kappa} \sim \sin(k r + \phi_\kappa) ,
\end{equation}
where $k$ is the radial momentum.
At the boundary of the box of radius $R_{\rm box}$ 
impose, for instance, that
\begin{equation}
\label{boundary}
k_{(0)}  R_{\rm box}  + \phi_\kappa^{(0)} = n 
\pi  \,\,\,\, n = 0, 1, 2, \ldots 
\end{equation}
where the index $0$ refers to the free modes.
From~(\ref{boundary}), 
the shift of energy level for fixed $\kappa$ is then given by 
\begin{equation}
E - E_0 \simeq {dE\over dk} (k - k_0) = 
- {1\over R_{\rm box}} \, {dE\over dk} \,(  \phi_\kappa -
 \phi^{0}_\kappa) = - {dE\over dk}\, {\delta_{\kappa}\over R_{\rm box}} 
\end{equation}
where $\delta_\kappa$ is the phase shift between the free and 
scattering eigenstates.
Given that  the number of modes per unit energy  is
\begin{equation}
{dn\over dE} = {1\over \pi} R_{\rm box} {dk\over dE} ,
\end{equation}
we finally get
\begin{equation}
\label{ss}
dn\,(E - E^{0}) = - {1\over \pi}\,  \delta_\kappa \,dE .
\end{equation}
Hence, Eq.~(\ref{discsum}) becomes~\cite{julian}
\begin{equation}
\label{sch3}
W = {1\over 2 \pi}\sum_{\kappa} \int_{0}^{\infty} dE\, \left[
	\delta_\kappa (E)+\delta_\kappa (-E) \right].
\end{equation}
Of course, just as in~(\ref{exp2}), the expression 
Eq.~(\ref{sch3}) is a formal, ultraviolet-divergent quantity that  must be 
renormalized.

Our strategy to address this issue is the following.
(Details may be found in Section~\ref{sec:scat},  where
we carry out an actual calculation.)
Consider the expansion of the vacuum energy 
in the external potential of Eq.~(\ref{exp2}). 
As noted above, only the term of second order in the external potential
is ultraviolet divergent.  The evaluation of this term involves the
vacuum polarization tensor $\Pi_{\mu\nu}$.  Its divergent
part can be absorbed using the standard renormalization procedure. 
The terms with~$k \geq 4$ in the expansion~(\ref{exp2})
are ultraviolet-convergent.  If $\alpha R \gg 1$ (with $R$
the effective range of the potential), however, 
the terms with $k \geq 4$ in~(\ref{exp2}) must be resummed in order
to get a sensible result. Using Eq.~(\ref{exp2}), this
seems like  a hopeless task.

Consider then the Taylor expansion of the phase shifts
$\delta_\kappa$ in powers of the external potential $\alpha 
R$. Only the first and second Born terms will lead to
ultraviolet
divergences  in the expression of the vacuum energy~(\ref{sch3}).
The first Born term is odd in $E$, and drops from the
calculation
if we integrate symmetrically over positive and negative energy
phase shifts. If we subtract the second
Born approximation from~(\ref{sch3}), the resulting, ``renormalized
vacuum energy'' is free of ultraviolet divergences.
The resulting quantity is equal to the resummed series
of Eq.~(\ref{exp2}) with $k\geq 4$ and is precisely 
the quantity which has been estimated in Ref.~\cite{fb}.

In Section~\ref{sec:scat} we compute the renormalized vacuum energy
for a spherical square well potential with depth $\alpha$
and radius $R$ using Eq.~(\ref{sch3}).
We numerically integrate the phase shifts over energy and subtract
the second Born term.  In principle we are required to evaluate
all of the terms in~(\ref{sch3}).  In practice, 
it is sufficient to investigate the convergence of the series
in order to obtain a numerical estimate for the vacuum energy.
As we will see, for large $\alpha R$ the
number of terms to be calculated grows linearly with  $\alpha R$.
Of course, we cannot expect to carry out such a calculation
for an actual neutron star, for which $\alpha R \sim 10^{12}$. 
This is not necessary, however. 
Because the neutrinos are massless, there is
just one dimensionless parameter, which is the effective expansion 
parameter $\alpha R$.  The question is then 
how the  ground state energy changes 
from the domain where the perturbative
expansion converges, $\alpha R \ll {\cal O}(1)$, to 
the domain where  the expansion presumably breaks
down because of the infrared
divergence, $\alpha R \gg {\cal O}(1)$. 

We have computed the renormalized vacuum energy for a number of
points in the range $0\leq \alpha R \leq 100$, 
including the point $\alpha R=20$.
According to Fischbach, {\em et al}~\cite{fb2}, 
the energy corresponding to this point should be on the  order of 
the mass of a neutron star, 
$W \sim 10^{30}\,{\rm kg} \sim 10^{66}$ eV. 
By way of comparison, we find  $ W  \simeq  - 2.3 $ keV.
The energy is well-behaved over the whole range of our calculation. 
Remarkably, the energy per unit volume of the potential  plotted
as a function of $\alpha R$ exhibits
a  crossover at  $\alpha R = \pi$. For $\alpha R > \pi$, the ground state
of the system carries a non-zero neutrino charge.

The neutrino charge of the vacuum in the potential $\alpha$ is
defined as
\begin{eqnarray}
\label{charge}
q  &=&  i \int d^3 x\, \mbox{\rm tr}
\left[S_F(x,x^\prime) - 
S_F^0(x,x^\prime)\right]_{x^\prime \rightarrow x}\\
&\equiv & - {1\over 2} \left\{\sum_{i>0} - \sum_{i<0} \right\}
\end{eqnarray}
using the notation introduced in~(\ref{discsum}).
Using Eqs.~(\ref{boundary})-(\ref{ss}) gives~(\ref{charge}) 
in terms of the phase shifts,
\begin{equation}
\label{qdelta}
q = -{1\over 2 \pi} \sum_\kappa 
\int_0^\infty dE\,\left \{{d 
\delta_\kappa(E)\over dE} - {d \delta_\kappa(-E)\over 
dE} \right\} .
\end{equation}
Just as was the case with the formal expressions for
the ground state energy, the above expression for the vacuum 
charge will need to be renormalized.  In this case this
is accomplished by subtracting out the first Born term.

The existence of a  neutrino condensate in the
ground state is to be expected~\cite{loeb}. Because the neutrinos 
are massless, neutrino-antineutrino pairs are {\em a priori} easily produced
by any non-zero gradient  of the potential.
Perhaps surprisingly, we have found that in our model it
takes a finite value, $\alpha R = \pi$, for this
to happen. 
For $\alpha R > \pi$, the ground state of our model
contains a net neutrino charge.
Note that the coupling term in~(\ref{efflag})
 is precisely analogous
to a constraint on the neutrino number with chemical potential 
$\mu=\alpha$~\cite{ken1}. In the large volume limit, we expect the charge of
the vacuum to be well approximated by 
that of a condensate with volume 
$V = 4 \pi R^3/3$\cite{abada},
\begin{equation}
\label{qcond}
 q_{\rm cond} =  \,V \, 
\int {d^3k\over (2 \pi)^3}  \theta(k -\alpha)
=  {2(\alpha R)^3\over 9 \pi} .
\end{equation}
Similarly, using the dispersion relation 
$\omega(k) = k - \alpha$ [which follows from (\ref{efflag}) in
the case of a homogeneous potential],
the renormalized ground state energy should approach~\footnote{In a
neutron star, the induced
neutrino charge density  is extremely small,  
$\sim  \,n_\nu\times 2\cdot 10^{-23} 
{\rm fm}^{-3}$ (for $\alpha \sim 20$ eV, and with $n_\nu$ the number
of massless neutrino species), compared to the neutron density, 
$\sim 0.4\; {\rm fm}^{-3}$.  Consequently the back-reaction of 
the neutrino condensate
on the neutron-induced  effective potential
is  totally negligible. Also, because the low energy 
neutrinos of the condensate
interact very weakly
with each other, higher orders effects are presumably very small.} 
\begin{equation}
\label{wcond}
W_{\rm cond} = V \, \int {d^3k\over (2 \pi)^3} 
(k - \alpha)  \theta(k -\alpha)
=  -  {\alpha^4 R^3 \over 18 \pi}
\end{equation}
That this is indeed the limiting  behaviour of the renormalized 
vacuum energy for large potentials is demonstrated in Section~\ref{sec:scat}.

\section{The neutrino ground state in  a  square well  potential}
\label{sec:scat}

\subsection{Derivation of the phase shift}

Both the charge and the energy of the neutrino ground
state in the presence of a background potential may
be expressed in terms of the scattering phase shifts.
Our starting point, then, is
the equation of motion following from the effective
Lagrangian given in Eq.~(\ref{efflag}),
\equation
	\left[\omega -i\vec{\sigma}\cdot\vec{\nabla} 
		+ \alpha(r)\right]\psi_L = 0,
\label{weyleqn}
\endequation
where $\psi_L$ is a two-component spinor, and where 
the potential is taken to be spherically symmetric.
The energy, $\omega$, may be positive or negative, with positive values
corresponding to particle solutions and negative values corresponding
to anti-particle solutions.

In order to solve for the scattering solutions we proceed, in the usual
way~\cite{itzykson,messiah}, by decomposing them into angular
momentum eigenstates,
\equation
	\psi_{L_{jm}} = f_{jm}(r)\Omega_{jm}^{(+)}+
		g_{jm}(r)\Omega_{jm}^{(-)},
	\label{decomposition}
\endequation
where
\begin{eqnarray}
	\Omega_{jm}^{(+)} & = & (2l+1)^{-\frac{1}{2}}
          \left(\begin{array}{c}
	  \left(l+m+\frac{1}{2}\right)^{\frac{1}{2}}Y_l^{m-\frac{1}{2}}\\
	  \left(l-m+\frac{1}{2}\right)^{\frac{1}{2}}Y_l^{m+\frac{1}{2}}
		\end{array} \right) ,\;\;\;\;\; l = j-\frac{1}{2} \\
	\Omega_{jm}^{(-)} & = & (2l^\prime+1)^{-\frac{1}{2}}
          \left(\begin{array}{c}
	  \left(l^\prime-m+\frac{1}{2}\right)^{\frac{1}{2}}
		Y_{l^\prime}^{m-\frac{1}{2}}\\
	  -\left(l^\prime+m+\frac{1}{2}\right)^{\frac{1}{2}}
		Y_{l^\prime}^{m+\frac{1}{2}}
		\end{array}\right),\;\;\;\;\; l^\prime = j+\frac{1}{2}>0 ,
\end{eqnarray}
and where we use the standard notation $Y_l^m$ for the
spherical harmonics.
The two-component spinors $\Omega_{jm}^{(\pm)}$ are eigenstates
of $J^2$ and $J_z$, with eigenvalues $j(j+1)$ 
($j$$=$$\frac{1}{2},\frac{3}{2},\ldots$) and $m$, respectively.
They are also eigenstates of $L^2$, with orbital angular momentum
$l$$=$$j-\frac{1}{2}$ for $\Omega_{jm}^{(+)}$
and $l^\prime$$=$$j+\frac{1}{2}$ for $\Omega_{jm}^{(-)}$.
In the following we will label the solutions by 
$l$$=0,1,2,\ldots$ instead of by $j$, although it is 
to be understood that the solutions themselves are eigenstates of 
the total, not the orbital, angular momentum.
With the above normalization for the angular momentum eigenstates
we have the useful relation
\equation
	\vec{\sigma}\cdot\hat{r} \; \Omega_{jm}^{(\pm)} = \Omega_{jm}^{(\mp)} .
\endequation
Inserting (\ref{decomposition}) into (\ref{weyleqn}), we then
obtain a set of coupled
first-order differential equations in $f_l$ and $g_l$,
\begin{eqnarray}
	\frac{df_l}{dr} -\frac{l}{r}f_l & = & 
		-i[\omega + \alpha(r)]g_l \label{ceq1}\\
	& &\nonumber \\
	\frac{dg_l}{dr} +\frac{l+2}{r}g_l & = & 
		-i[\omega + \alpha(r)]f_l \label{ceq2}.
\end{eqnarray}

As noted above, we will use a spherical square well potential
in order to simplify our calculations.  This choice has the benefit
that the expression for the phase shift may be written out explicitly
in terms of simple functions.  We thus take
\equation
	\alpha(r) = \alpha\theta(R-r), \;\;\;\;\; r>0.
\endequation
The solutions of Eqs.~(\ref{ceq1}) and 
(\ref{ceq2}) are then simply given by 
spherical Bessel functions.  Requiring that
the solutions be regular at the origin, and dropping an over-all 
normalization factor, we find
\begin{eqnarray}
	f_l & = & \left\{\begin{array}{ll}
		j_l(|\omega+\alpha|r),\;\;\;\;\; & 0<r<R \\
		& \\
		B \left[j_l(|\omega|r) \cos\delta_l+ 
			n_l(|\omega|r)\sin\delta_l\right],\;\;\;\;\; & r>R
 		\end{array}\right. \label{eq:fl}\\
	& & \nonumber\\
	g_l & = & \left\{\begin{array}{ll}
		-i \varepsilon(\omega+\alpha)j_{l+1}
			(|\omega+\alpha|r),\;\;\;\;\; & 0<r<R \\
		& \\
		-i \varepsilon(\omega)B\left[j_{l+1}(|\omega|r) \cos\delta_l+ 
			n_{l+1}(|\omega|r)\sin\delta_l\right],\;\;\;\;\; 
			& r>R , \label{eq:gl}
		\end{array}\right.
\end{eqnarray}
where $B$ and $\delta_l$ are fixed by matching
$f_l$ and $g_l$ at $r$$=$$R$ (note that the derivatives of 
$f_l$ and $g_l$ are in general discontinuous across the boundary) and
where
\equation	
	\varepsilon(x) \equiv \frac{x}{|x|} .
\endequation
Our conventions for the spherical Bessel functions are as follows:
\equation
	j_l(\rho) = \left(\frac{\pi}{2\rho}\right)^{\frac{1}{2}}
		J_{l+\frac{1}{2}}(\rho),\;\;\;\;\;\;
	n_l(\rho) = (-1)^l\left(\frac{\pi}{2\rho}\right)^{\frac{1}{2}}
		J_{-l-\frac{1}{2}}(\rho).
\endequation
Performing the matching at the boundary, we finally
arrive at the desired expression for the phase shift
\equation
	\tan\delta_l(\omega) = \frac{\varepsilon(\omega)j_l(|\omega+\alpha|R)
		j_{l+1}(|\omega|R)-\varepsilon(\omega+\alpha)j_l(|\omega|R)
		j_{l+1}(|\omega+\alpha|R)}{
		\varepsilon(\omega+\alpha)n_l(|\omega|R)
		j_{l+1}(|\omega+\alpha|R)-\varepsilon(\omega)
		j_l(|\omega+\alpha|R)n_{l+1}(|\omega|R)} .
	\label{phaseshift}
\endequation
We may now use this expression in order to calculate the energy
and charge of the vacuum.
	
\subsection{Energy and charge of the vacuum}

The formal expressions relating the energy and charge of the vacuum
to the scattering phase shifts are given in (\ref{sch3}) and (\ref{qdelta}).
The spherical symmetry in our problem implies that the sum over
$\kappa$ becomes a sum over $l$.  We may thus rewrite these
expressions as
\begin{eqnarray}
	W & = & \frac{1}{2\pi} \sum_{l=0}^{\infty} (2l+2) \int_{0}^\infty
		d\omega \; \left[\delta_l(\omega)+\delta_l(-\omega)
		\right] \label{wformal}\\
	& & \nonumber \\
	q & = & -\frac{1}{2\pi}\sum_{l=0}^{\infty} (2l+2) \int_{0}^\infty
		d\omega \;
		\left[\frac{\partial}{\partial\omega}\delta_l(\omega)-
		\frac{\partial}{\partial\omega}\delta_l(-\omega)\right] .
	\label{qformal}
\end{eqnarray}
The factor $(2l+2)\equiv (2 j +1)$ is the degeneracy factor for a given
energy $\omega$ and total angular momentum $j$.

As we have discussed in Section~\ref{sec:rev}, the above formal expressions
are in need of renormalization.  Since our model is renormalizable,
the divergences are confined to the first few Born terms.
The procedure to be followed in handling these 
divergences is then as follows: (i) Taylor-expand the formal expressions 
for the energy and the charge in order to isolate the divergences; (ii)
subtract out the divergent terms and ``resum'' the Taylor expansion; (iii)
renormalize the divergent terms in the usual way and add them back in.
The resulting renormalized expressions are then finite and correspond
to the actual energy and charge of the ground state of the system.

The first step in renormalizing the energy and the charge is to 
perform a Taylor expansion of the phase shift in $\alpha R$.  Let us
then define
\equation
	\delta_l(\omega;\alpha R) = (\alpha R) \delta^{(1)}_l(\omega) + 
		(\alpha R)^2 \delta^{(2)}_l(\omega) + 
		(\alpha R)^3 \delta^{(3)}_l(\omega) + \ldots,
	\label{delborn}
\endequation
where
\equation
	\delta^{(n)}_l(\omega) \equiv \frac{1}{n!}
		\left. \frac{\partial^n}{\partial z^n} \delta_l(\omega;z)
		\right|_{z\to 0} .
\endequation
The explicit expressions for the first two Born terms are given by
\begin{eqnarray}
	\delta^{(1)}_l(\omega) & = & \varepsilon(\omega) \beta^2\left\{
		\left[j_l(\beta)\right]^2+\left[j_{l+1}(\beta)\right]^2-
		\frac{2(l+1)}{\beta}j_l(\beta)j_{l+1}(\beta)\right\}, 
	\label{del1def} \\
	& & \nonumber\\
	\delta^{(2)}_l(\omega) & = & -\beta \left\{ \left[
		\left[j_l(\beta)\right]^2+\left[j_{l+1}(\beta)\right]^2
		-\frac{3(l+1)}{\beta}j_l(\beta)j_{l+1}(\beta)\right] \right.
		\nonumber \\
	& & \;\;\;\; -\beta^3\left[n_l(\beta)j_l(\beta) + 
		n_{l+1}(\beta)j_{l+1}(\beta) - \frac{1}{\beta}\left(
		l j_l(\beta)n_{l+1}(\beta) + (l+2)j_{l+1}(\beta)n_l(\beta)
		\right)\right] \nonumber \\
	& & \;\;\;\; \times \left. \left[\left[j_l(\beta)\right]^2+
		\left[j_{l+1}(\beta)\right]^2
                -\frac{2(l+1)}{\beta}j_l(\beta)j_{l+1}(\beta)\right]\right\} ,
	\label{born2}
\end{eqnarray}
where we have defined $\beta$$\equiv$$|\omega|R$.  It is easy to convince
oneself that the even (odd) Born terms are even (odd) in $\omega$.

The formal expression for the energy is rendered finite
by subtracting out the second Born term.  (Since the first
Born term is odd in $\omega$, it drops out of the calculation
if we integrate symmetrically over positive and negative 
energies.\footnote{Note that all terms in the Born expansion
which are odd in $\alpha$ are also odd in $\omega$ and thus do not
contribute to the energy.  See the footnote following Eq.~(\ref{exp2}).})
To compute the second Born term, we may use the expression
for the $k=2$ term in the expansion (\ref{exp2}).  Using the standard
prescription, we may then regularize and renormalize this term.
The renormalized expression for the energy is then 
given by
\equation
        W_{\rm ren} = W_{\rm ren}^{(2)} + W^{(4+)},
	\label{wren2}
\endequation
where
\equation
	W^{(4+)} = \frac{1}{2\pi}\sum_{l=0}^{\infty} 
		(2l+2) \int_0^\infty
                d\omega \; \left[\delta_l(\omega)+\delta_l(-\omega)-
                2(\alpha R)^2 \delta^{(2)}_l(\omega)\right] ,
	\label{e4plus}
\endequation
and where $W_{\rm ren}^{(2)}$ denotes the renormalized
vacuum polarization contribution.
$W^{(4+)}$ corresponds to the resummed terms in the 
expansion (\ref{exp2}) with
$k\geq 4$.

The formal expression for the charge in (\ref{qformal}) is 
divergent because the phase shifts tend to a constant as 
$\omega\to\infty$~\cite{mani},
\equation
	\lim_{\omega\to\pm\infty} \delta_l(\omega) = 
		\pm \int_0^{\infty} \alpha(r) dr = \pm \alpha R .
	\label{dellim}
\endequation
This divergence is removed by subtracting out the first Born
term, which, as may easily be verified from (\ref{del1def}),
satisfies
\equation
	\lim_{\omega\to\pm\infty} \delta^{(1)}_l(\omega) = \pm 1,\;\;\;\;\;
		\delta^{(1)}_l(0) = 0 .
	\label{del1lim}
\endequation
Once regularized, this term gives a vanishing contribution to
the renormalized charge (the charge is actually zero to all
orders in perturbation theory), so that the final expression for
the renormalized charge is given by
\equation
	q_{\rm ren}  =  \frac{1}{2\pi}\sum_{l=0}^{\infty} 
		(2l+2)\left[\delta_l(0^+)-\delta_l(0^-)\right] .
	\label{qren2}
\endequation
The problem of calculating the charge of the vacuum then simply reduces
to that of evaluating the particle and anti-particle phase shifts at the
origin.

It is useful to note that for small $\alpha R$ the Taylor
expansion of the phase shift given in (\ref{delborn}) may be used
to obtain a perturbative expansion for $W^{(4+)}$.  For future reference,
let us write this perturbative expansion as follows:
\equation
	W^{(4+)} = \frac{1}{2\pi R}\sum_{n=4,6,8,\ldots}
		(\alpha R)^n\sum_{l=0}^{\infty} 
		(2l+2) I_l^{(n)},\;\;\;\;\;\;\;\;\;
		\mbox{\rm (small $\alpha R$)}
	\label{e4pluspert}
\endequation
where $I_l^{(n)}$ denotes the integral over $\omega$ of the $n^{\rm th}$
Born term,
\equation
	I_l^{(n)} \equiv 2 \int_0^\infty d(\omega R)\, 
		\delta^{(n)}_l(\omega) .
	\label{intdel}
\endequation
This expansion is expected to be reliable for small $\alpha R$.
For large $\alpha R$, of course, the perturbative expansion breaks down.

The main goal in the remainder of this paper will be to study the 
behaviour of the renormalized energy and charge as a function of 
$\alpha R$.  As we shall see, for fixed $\alpha$ both of these 
quantities scale like the volume of the potential region 
as the volume gets large.
Before studying these quantities in detail, however, let us first consider
the contribution to the total energy due to the vacuum polarization,
$W^{(2)}_{\rm ren}$.  Since $W^{(2)}_{\rm ren}$ scales roughly like
the surface area of the potential region
(for fixed $\alpha$), its contribution to 
the total energy is eventually overwhelmed by that due
to $W^{(4+)}$.  Nevertheless, the handling of this term is somewhat
subtle, so it is important to consider it carefully.
Once we have studied $W^{(2)}_{\rm ren}$, we will dispense with it
altogether and consider only the contribution due to $W^{(4+)}$ when
discussing the renormalized energy.

\subsection{Vacuum polarization}

We now turn to the calculation of $W^{(2)}_{\rm ren}$. Consider the
$k=2$ term in the expansion (\ref{exp2}),
\begin{equation}
\label{e2}
W^{(2)} =  {1\over 4 \pi i} \int d\omega\, \int d^3x_1 d^3x_2\,\,
\mbox{\rm tr}\left [ \alpha(\vec x_1) G_0(\vec x_1,\vec x_2) \alpha(\vec x_2) 
G_0(\vec x_2, \vec x_1)\right] .
\end{equation}
Using
\begin{equation}
G_0(\vec x) = -  \int {d^3 k\over (2 \pi)^3}\, {e^{i {\vec k\cdot \vec x}}\over
\omega - {\vec \sigma \cdot \vec k}} ,
\end{equation}
leads to
\begin{equation}
W^{(2)} = - {1\over 2 i} \int d^3x_1 d^3x_2 \,\int {d^3{p}\over
(2 \pi)^3} \alpha({\vec x_1}) \alpha({\vec x_2}) \Pi_{00}({\vec p})
\, e^{i {\vec p \cdot (\vec x_1 - \vec x_2)}} ,
\end{equation}
where
\begin{equation}
\label{vacpol}
\Pi_{\mu\nu}(p) = - {i\over 24 \pi^2} \left(g_{\mu\nu} - {p_\mu p_\nu\over
p^2}\right)\, p^2\, \left\{{2\over \epsilon} - \gamma + \log 4
\pi + {5\over 3} - \log\left({-p^2\over \mu^2}\right) \right \}
\end{equation}
is the familiar one-loop vacuum polarization tensor, which we have 
calculated using dimensional
regularization. In Eq.~(\ref{vacpol}), $p$ is the 
4-momentum in Minkowski space and $\mu$ is the renormalization scale.
The divergent term, $\propto 1/\epsilon$,
(plus  an arbitrary constant term)
can be  absorbed as usual by redefining the parameters of the theory,
$\alpha \rightarrow \alpha(\mu)$. Note that, because the polarization tensor
is proportional to $p^2$, the renormalized vacuum polarization term
vanishes if the potential is 
homogeneous~\cite{abada}. 
Introducing 
\begin{equation}
\tilde \alpha(\vert\vec p\vert ) = \alpha \int d^3 x \, 
	e^{i {\vec p \cdot \vec x}}
\theta(R - r) = { 4\pi \alpha R^2\over \vert \vec p\vert } \, 
	j_1(\vert \vec p\vert R) ,
\end{equation}
with $j_1(x) = \sin(x)/x^2 - \cos(x)/x$,  gives finally
\begin{equation}
\label{rene}
W^{(2)}_{\rm ren} = {\alpha^2 R\over 6 \pi^2} \, \int_0^\infty dz\, z^2 \,
 j_1^2(z) \,\log\left (
{z^2\over (\mu R)^2}\right) .
\end{equation}
As it stands, the renormalized expression in~(\ref{rene}) is still
a linearly divergent quantity. We have chosen to work with a 
spherical square well potential and, because of the
presence of the sharp edge, arbitrarily high radial momenta contribute to
the integral.  Any smoother potential would lead to a finite result.
At the level of the effective theory, there is
a natural ultimate cut-off provided by the size of the neutron,
$\Lambda \sim 0.2$ GeV.  (This also suggests the choice $\mu = \Lambda$.)
Inserting the cut-off, we get that the
vacuum polarization term~(\ref{rene}) scales roughly
like the area of the potential, $W_{\rm ren}^{(2)} \sim
\alpha^2\, R^2 \,\Lambda$. Consequently, 
 $W^{(2)}_{\rm ren}/R^3$  contributes a vanishing amount to 
the  energy density as the size of the system increases.
Let us emphasize that the remainder of our calculation 
is completely insensitive to the presence of this cut-off since 
the higher order terms in (\ref{exp2}) are ultra-violet convergent.

\subsection{Numerical evaluation of the renormalized energy and charge}

We turn now to a numerical evaluation of the renormalized energy and 
charge.  From now on we will ignore the vacuum
polarization contribution to the energy.  We will thus loosely refer
to $W^{(4+)}$ as the renormalized energy.

\subsubsection{Small values of $\alpha R$}

\begin{figure}[hbt]
\centerline{\epsfig{figure=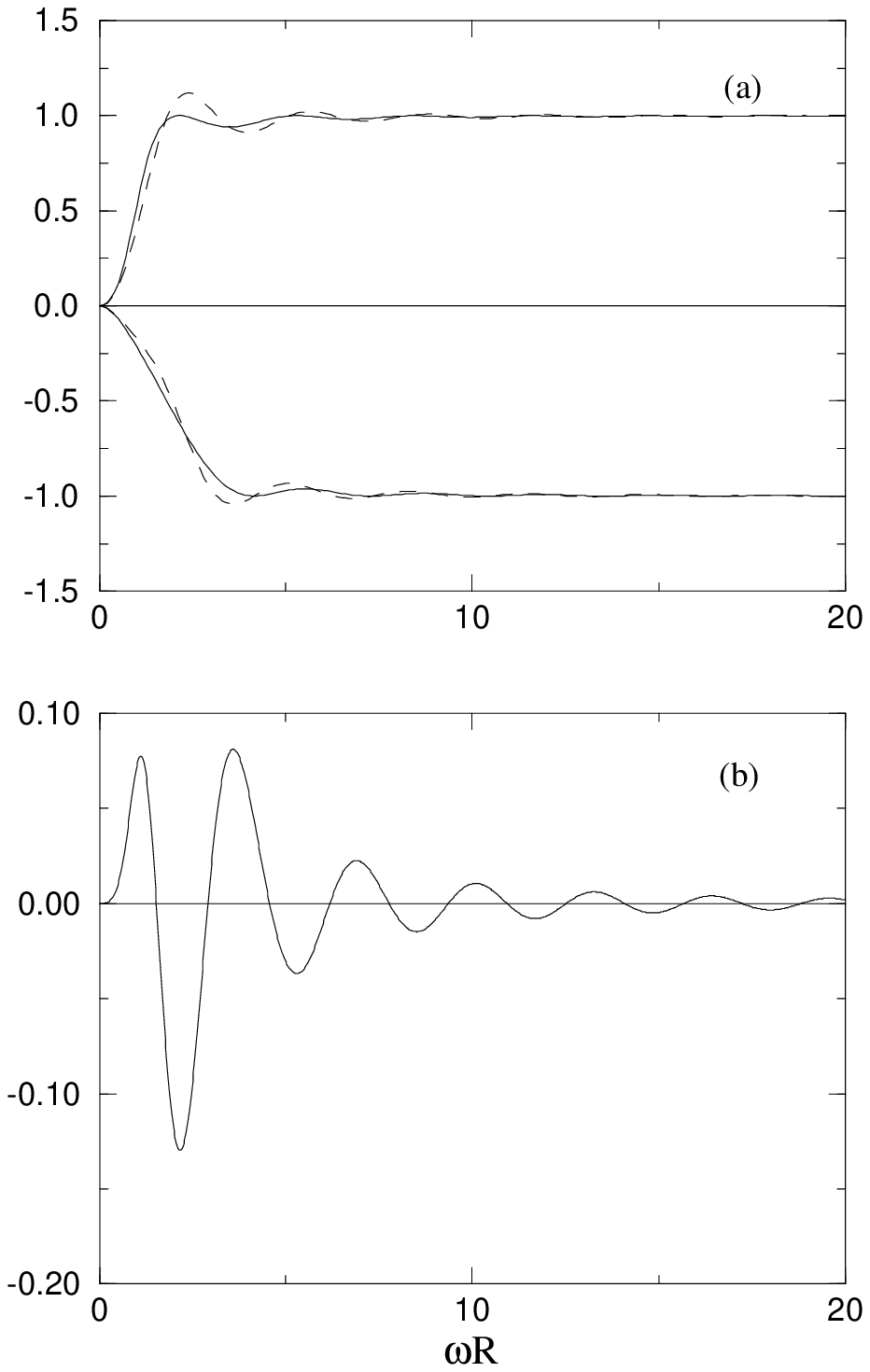,height=180mm}}
\caption{Plots of the phase shift for $\alpha R = 1$ and $l=0$.
(a) The upper (lower) solid curves give the particle (antiparticle)
phase shifts, while the dashed curves give the sums of the first and
second Born terms.  (b) This curve shows the sum of the particle and
antiparticle phase shifts with twice the second Born term subtracted.}
\label{fig:ar1}
\end{figure}

Let us begin our study of the charge and energy 
of the ground state by considering rather modest values of $\alpha R$.
Fig.~\ref{fig:ar1}(a) shows plots of
the particle and antiparticle phase shifts (solid curves) 
as functions of $\omega$ for $l=0$ in the case
that $\alpha R=1$.  Also shown 
are the sums of the first and second Born approximations to
the phase shifts (dashed curves).
Note that the particle (antiparticle)
phase shifts approach $+(-)\alpha R$ as $\omega\to\infty$.
Furthermore, the phase shifts go to zero at the origin so
that, according to Eq.~(\ref{qren2}), the vacuum charge is zero.\footnote{
It may at first seem surprising that the charge vanishes for $\alpha R = 1$. 
After all, the neutrinos are massless and it would seem that 
any gradient of the potential could  produce
neutrino-antineutrino pairs. However, this argument neglects
the effect of the neutrino's spin. To see this, consider the system
of first-order differential equations for $f_l$ and
$g_l$,~(\ref{ceq1}) and 
(\ref{ceq2}). 
Writing these as two decoupled second-order Schr\"odinger equations shows that
 the $g_l$ component
always  has a  repulsive centrifugal barrier,
$$
{d^2 f_l\over dr^2} + {2\over r} {d f_l\over dr} - {l(l+1)\over
r^2}f_l + 
(\omega +\alpha)^2 f_l = 0
$$
$$
{d^2 g_l\over dr^2} + {2\over r} {d g_l\over dr} - {(l+1)(l+2)\over
r^2}g_l + 
(\omega +\alpha)^2 g_l = 0
$$
Thus, even for the $l=0$ mode, it takes a finite $\alpha R$ to create
 a charge
 in the vacuum.
}
The plots for higher values of $l$ are qualitatively similar.

Fig.~\ref{fig:ar1}(b) shows a plot -- again for $\alpha R = 1$ and
$l=0$ -- of the
sum of the particle and anti-particle
phase shifts with the second Born term subtracted.  This is
the quantity which must be integrated (and summed over $l$) in order
to obtain the renormalized energy in Eq.~(\ref{e4plus}).

For small enough values of $\alpha R$, one expects the 
energy to be well-described by the leading term in the 
perturbative expansion defined in (\ref{e4pluspert}).
Our numerical results show that this is indeed the case
for $\alpha R\lsim 1/2$, which actually provides a non-trivial
check on our results.  The advantage of
using the perturbative expansion to evaluate the renormalized
energy for small $\alpha R$ is that the integrals over $\omega$
may be done exactly in that case,\footnote{We do not include the explicit form
of $\delta^{(4)}_l(\omega)$ here because it is rather unwieldy.
As was the case for $\delta^{(2)}_l(\omega)$, however, the expression
for $\delta^{(4)}_l(\omega)$ may be expressed in terms of spherical
Bessel functions.  Thus the integrals may
be done analytically (albeit with the aid of a computer).}
whereas those for the exact expression, (\ref{e4plus}),
must be done numerically.  Furthermore, in the case of the
exact expression, the numerical integrations 
over $\omega$ become increasingly difficult as $\alpha R\to0$.

\begin{figure}[hbt]
\centerline{\epsfig{figure=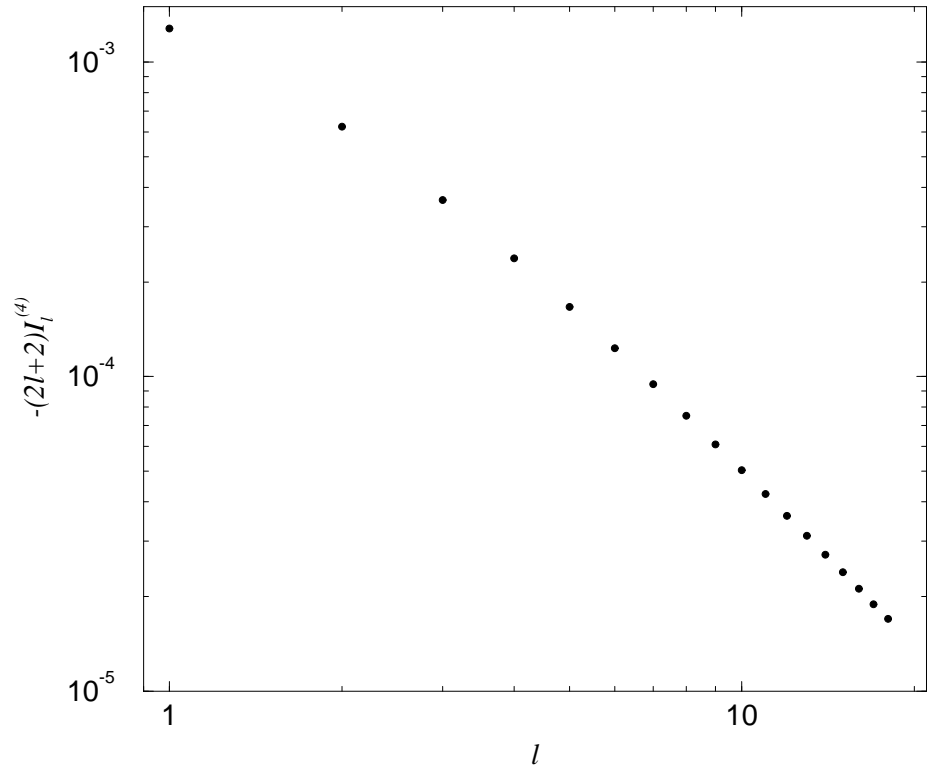,height=120mm}}
\caption{Plot of $-(2l+2)I_l^{(4)}$ vs. $l$.}
\label{fig:del4}
\end{figure}

The leading term in the perturbative expansion is at fourth
order in $\alpha R$.  
Fig.~\ref{fig:del4} shows a log-log plot of $(-1)\times(2l+2)\times I_l^{(4)}$
as a function of $l$.  (We have not plotted the point corresponding to
$l=0$ for obvious reasons.)  The sum over $l$ of this quantity is directly
proportional to the renormalized energy in the perturbative region.
As is clear from the figure, this quantity has roughly a power-law
fall-off as $l\to\infty$.  Using the slope of the curve 
in Fig.~\ref{fig:del4}, we find that $(2l+2)\times I_l^{(4)}\propto l^{-1.9}$,
for large $l$.  This allows us to obtain an approximate answer
for the sum over $l$,
\equation
	\sum_{l=0}^\infty (2l+2) I_l^{(4)} = -0.00721\pm 0.00005 ,
\endequation
where the quoted uncertainty is a rather conservative estimate 
of the error incurred in the extrapolation to large $l$.
We are thus able to compute the leading perturbative contribution to
$W^{(4+)}$, which is given by
\equation
	W^{(4+)} = -0.00115 (\alpha R)^4/R + \ldots
	\label{e4plustrunc}
\endequation

\subsubsection{Larger values of $\alpha R$}

\begin{figure}[hbt]
\centerline{\epsfig{figure=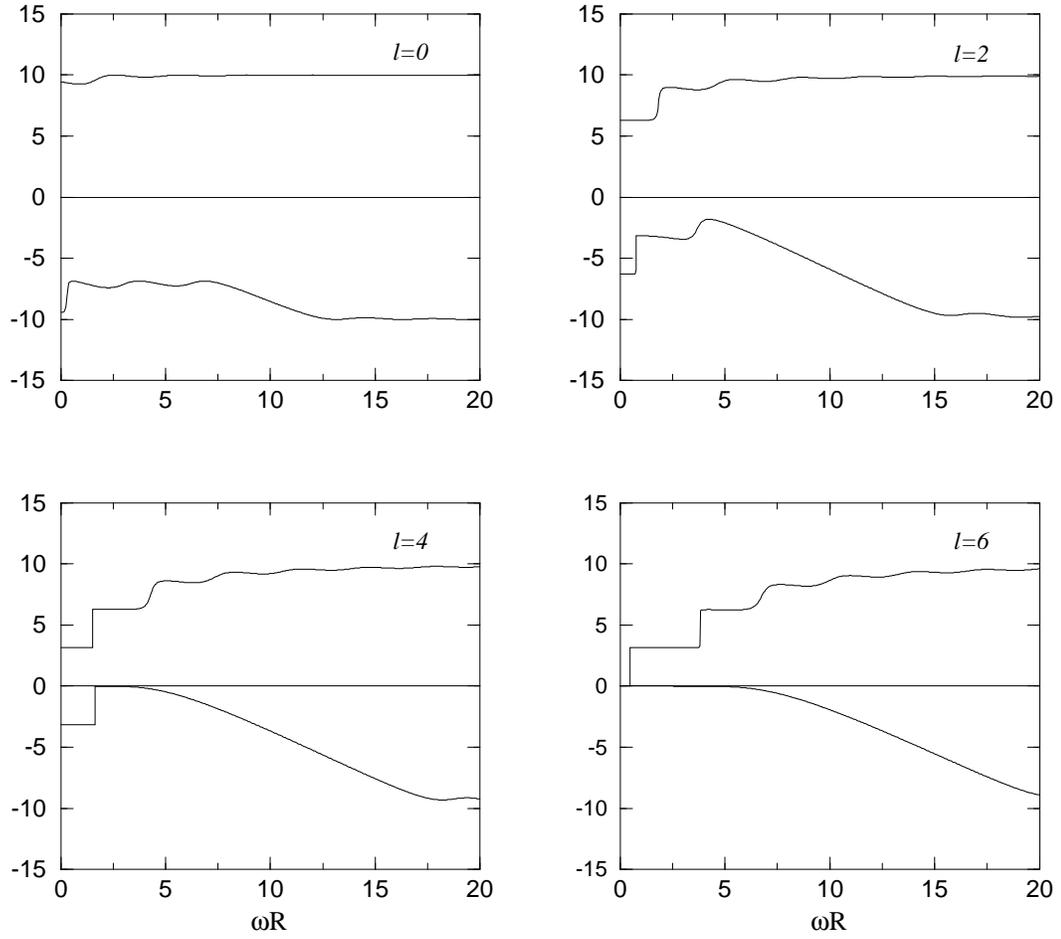,height=150mm}}
\caption{Plots of the particle and antiparticle phase shifts
for $\alpha R=10$ and $l=0,2,4$ and $6$.  In each case the upper
curve is the particle phase shift and the lower curve is the 
antiparticle phase shift.}
\label{fig:ar10}
\end{figure}

As $\alpha R$ is increased, one finds that there is a critical
point beyond which it becomes energetically favourable 
for the neutrinos to condense; thus, for 
$\alpha R$$>$$\left.\alpha R\right|_{\rm crit}$, the ground state of the
system contains a neutrino condensate.
The presence of the condensate has a very interesting 
effect on the renormalized energy.  
Before calculating the energy, however, let us first examine the
phase shifts for large $\alpha R$.  Fig.~\ref{fig:ar10} shows 
plots of the particle and antiparticle phase shift for $\alpha R = 10$,
with $l=0,2,4$ and 6.  These plots are qualitatively very different
from those for small $\alpha R$ [c.f. Fig.~\ref{fig:ar1}(a)] in that
the phase shift in the strong-potential case exhibits resonances.
Furthermore, for $l$ less than some critical value the phase shifts
do not vanish at the origin, as they do in
the weak-potential case.\footnote{Note, however, that 
Levinson's theorem is still satisfied: since there are no bound states,
the sum of the particle and anti-particle phase shifts at the origin is
zero~\cite{mani}.}  Thus, according to Eq.~(\ref{qren2}), the ground
state is charged:
\equation
	q_{\rm ren}(\alpha R = 10) = \sum_{l=0}^{l_{\rm max}} q_l
			= 56 .
\endequation
Note that the sum over $l$ for the charge always truncates at 
some $l_{\rm max}$.  For $\alpha R=10$, $l_{\rm max}=5$.

The charge of the ground state as a function of $\alpha R$ is intimately
tied to the resonance structure of the particle and antiparticle 
scattering solutions.  It is well-known in non-relativistic scattering
theory that as a potential well is deepened, the resonances of the 
scattering solutions migrate to the origin of the complex $k$-plane
and, at some critical strength of the potential, join the imaginary
axis and become true bound states.  As the potential is made deeper
and deeper, the bound states move to lower and lower energies (i.e.,
they become more and more tightly bound).  The situation is slightly different
in the relativistic scattering theory for fermions.  Consider 
first the case in which the fermion has a non-zero mass.  
In that case, the resonances of the positive energy (particle)
scattering solutions can migrate to the imaginary 
axis in the $k$-plane and become true
bound states, just as in the non-relativistic case.  As the potential
is deepened still more, however, the bound states eventually
turn around and re-enter the continuum, this time as resonances
for the negative energy (antiparticle) scattering solutions~\cite{raf}.
The presence of resonances in the negative energy scattering solutions
signals that the ground state has become charged.  
The situation is similar in the massless case, except that
in this case the positive energy resonances
move to the origin and pass immediately to the negative energy continuum
without ever making true bound states.  

In our simple model, it is possible to calculate precisely
the critical values of the potential at which 
the charge of the ground state changes discontinuously.
The solutions of the ``zero-energy resonances'' of the system
are normalizable solutions of the Weyl equation.\footnote{These are the 
``would-be'' bound states of the system which, in a vector-like
theory with a massive fermion, would have migrated onto the 
imaginary axis in the $k$-plane and become bound states.}
Setting $\omega$ to zero in Eqs.~(\ref{ceq1}) and (\ref{ceq2}),
we find that $f_l$ and $g_l$ decouple for $r>R$, giving
\begin{eqnarray}
	f_l & \propto & r^l \\
	g_l & \propto & r^{-(l+2)} ,
\end{eqnarray}
for $r>R$.  The normalizable solutions,
then, are those which kill $f_l$ in the asymptotic region.  After a 
few lines of algebra, one arrives at the following simple condition
for the critical parameters at which the charge of the vacuum changes:
\equation
	j_l(\alpha R) = 0 .
	\label{eq:jlzero}
\endequation
In particular, then, the vacuum goes from being uncharged to having
charge equal to $2$ when $\alpha R = \pi$.

\begin{figure}[hbt]
\centerline{\epsfig{figure=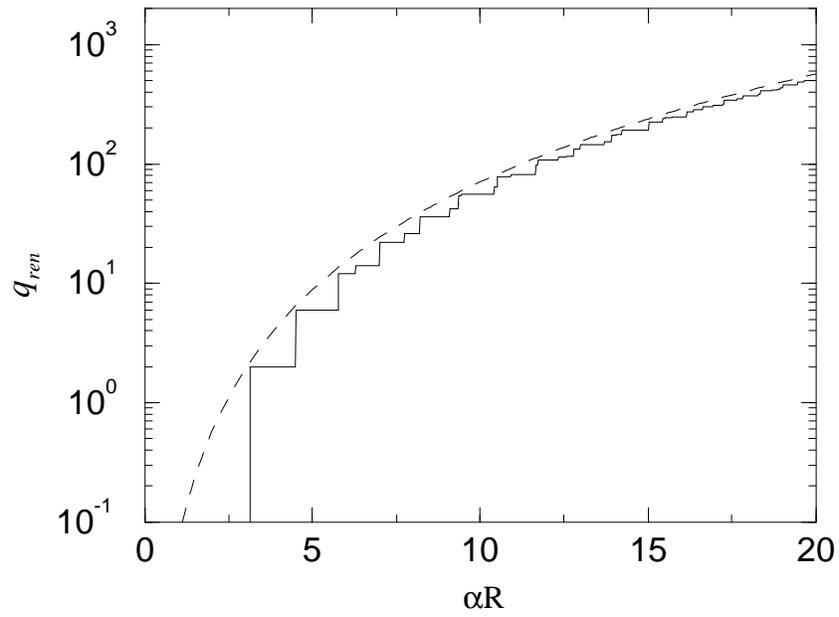,height=180mm}}
\caption{Plot of the charge as a function of $\alpha R$.
The solid curve gives the exact charge, which has periodic jumps,
and the dashed curve gives the charge expected for a condensate
in the large-volume limit.}
\label{fig:qren}
\end{figure}

Using the relation (\ref{eq:jlzero}), it is
straightforward to calculate the charge of the ground state for 
any value of $\alpha R$: for each $l$, one simply needs to count
the number of zeros of $j_l(\rho)$ with $\rho<\alpha R$.  
Fig.~\ref{fig:qren} shows a plot of the charge of the ground
state as a function of $\alpha R$.  It is clear from the solid curve
in this plot that the charge does indeed change discontinuously as 
$\alpha R$ is increased.  The dashed curve in this plot
shows the charge which one obtains in the large-volume limit
for a system with chemical potential $\mu = \alpha$, Eq.~(\ref{qcond}).
As $\alpha R$ is increased, we find that our exact result 
tends to this value.

\begin{figure}[hbt]
\centerline{\epsfig{figure=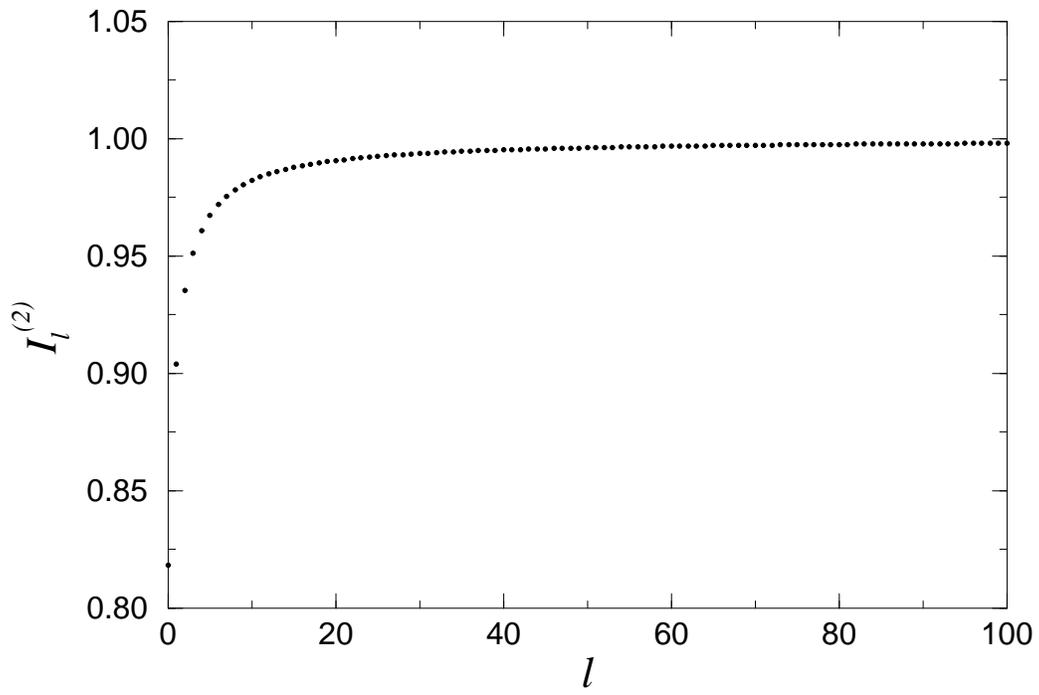,height=180mm}}
\caption{Plot of $I_l^{(2)}$ vs. $l$.}
\label{fig:del2}
\end{figure}

Let us now calculate the renormalized energy,  $W^{(4+)}$, for 
$\alpha R>\pi$.  Let us define
\equation
	W^{(4+)}\equiv \sum_{l=0}^\infty W^{(4+)}_l,
	\label{e4plus2}
\endequation
where
\begin{eqnarray}
	W^{(4+)}_l & = & \frac{1}{2\pi}(2l+2)\int_0^\infty
                d\omega \; \left[\delta_l(\omega)+\delta_l(-\omega)-
                2(\alpha R)^2 \delta^{(2)}_l(\omega)\right] \\
	& = & \frac{1}{2\pi}(2l+2)\int_0^\infty
                d\omega \; \left[\delta_l(\omega)+\delta_l(-\omega)\right]
		-\frac{(\alpha R)^2}{2\pi R} (2l+2)I_l^{(2)}
	\label{e4plusl}
\end{eqnarray}
In the second line we have explicitly separated out the term corresponding
to the integral of the second Born term, 
$I_l^{(2)}$, since it may be calculated 
analytically. (See Fig.~\ref{fig:del2}.)
We stress that for a given $l$ both terms in (\ref{e4plusl})
are finite.  

\begin{figure}[hbt]
\centerline{\epsfig{figure=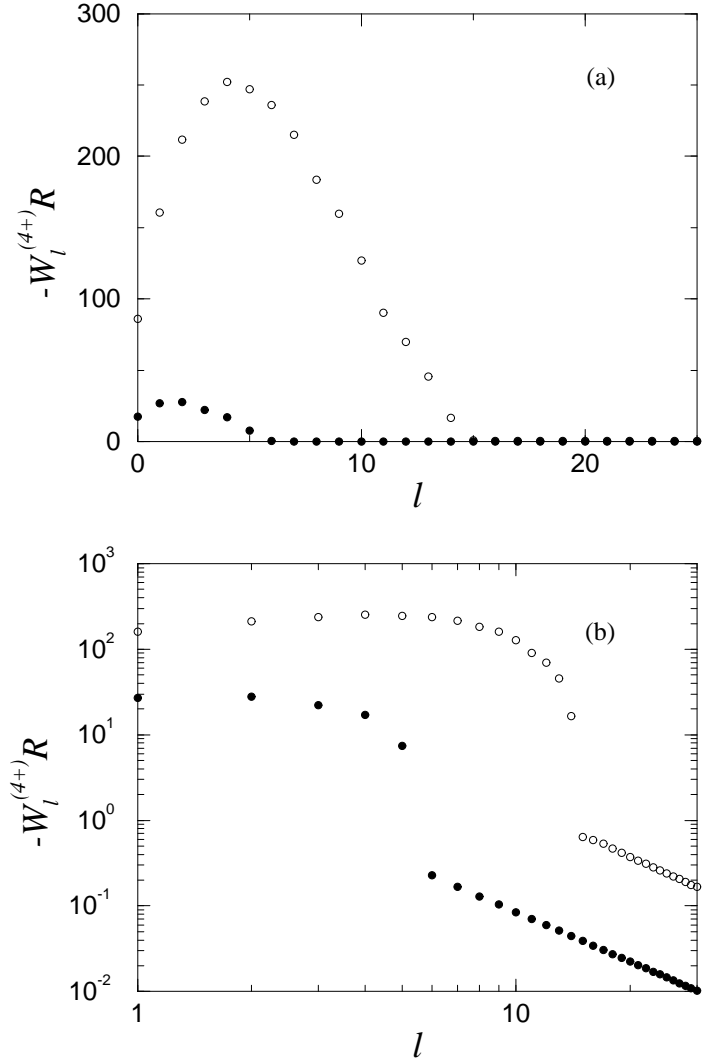,height=150mm}}
\caption{Plots of $-W_l^{(4+)}R$ vs. $l$ for $\alpha R = 10$ (solid dots)
and $\alpha R = 20$ (open circles).  The slopes of the tails in
the log-log plots are approximately $-2$ in each case.}
\label{fig:enl}
\end{figure}

Fig.~\ref{fig:enl} shows two plots of the dimensionless
combination $W_l^{(4+)} R$ as a function of $l$
for two strong potentials.
In each plot the solid dots correspond to $\alpha R=10$
and the open circles correspond to $\alpha R=20$.
These plots differ  qualitatively from the analogous
plots in the case of weak potentials. In the present case,
 the energy contains
a large ``bump'' for $l \leq l_{\rm crit}$
(see Fig.~\ref{fig:del4} for comparison).  For $l>l_{\rm crit}$,
the energy has a power-law fall-off 
which is similar to the large-$l$ behaviour found in the weak-potential case.
 From the slope of the tail in the log-log plot, one
sees that for large $l$ the tail goes roughly as $l^{-2}$.~\footnote{Note
 that a very precise determination (to better than one part in $10^5$)
 of the two terms contributing to $W_l^{(4+)}$ in Eq.~(\ref{e4plusl}) is
required in order to see that the tail in Fig.~\ref{fig:enl}(b) is
actually converging.} The presence of the bump
in the plot of $W_l^{(4+)}$ is directly related to the presence 
of a 
condensate.  In all cases which we have studied, the value of $l$
at which the energy suddenly drops is  precisely the same 
value of $l$ for which there is no more
contribution to the vacuum charge; {\em i.e.} $l_{\rm crit} =
l_{\rm max} \sim \alpha R $.

\begin{figure}[hbt]
\centerline{\epsfig{figure=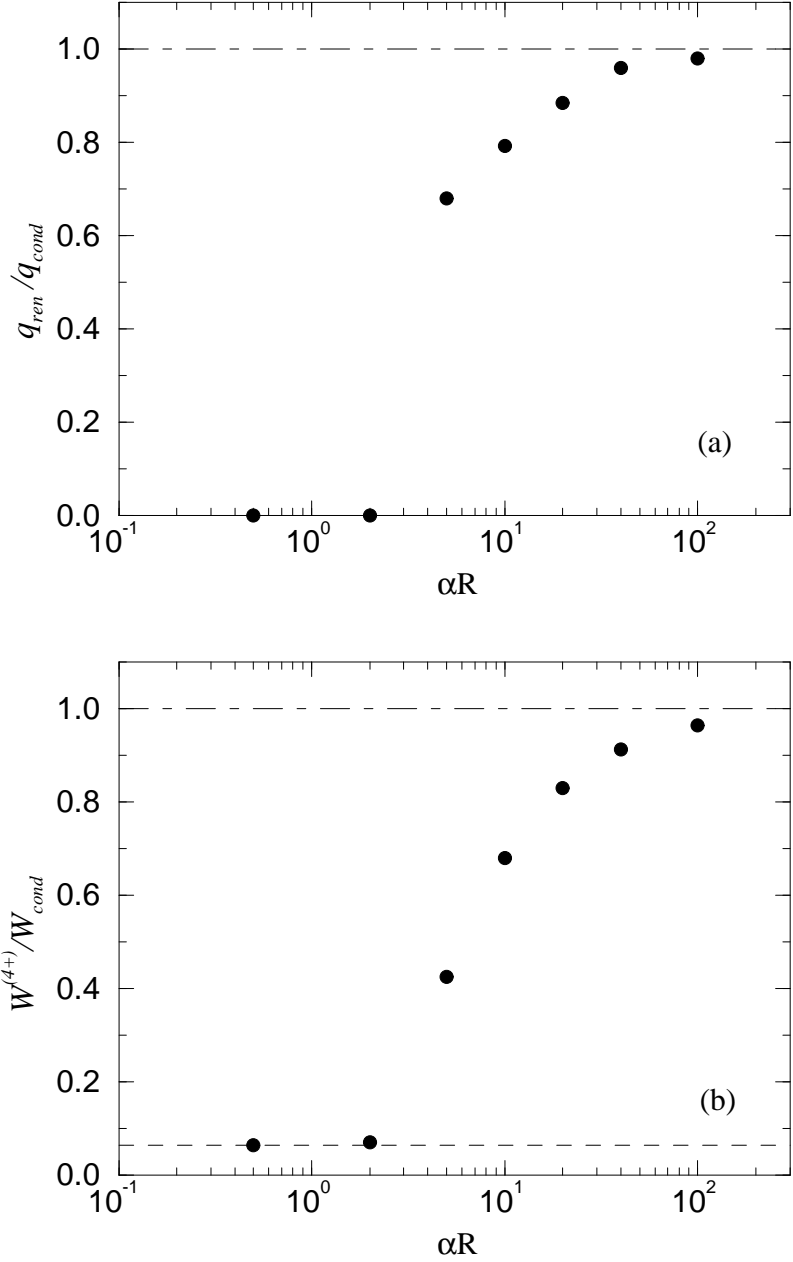,height=180mm}}
\caption{Plots of (a) the charge and (b) the 
energy as a function of $\alpha R$.  The dots give the results of our
exact (non-perturbative) calculations.  Both the charge and the
energy are normalized to the contribution expected from a 
condensate in the large-volume limit.  The lower dashed curve
in (b) gives the value obtained in the limit of small
$\alpha R$ (see Eq.~(\ref{e4plustrunc})).}
\label{fig:chargeen}
\end{figure}

The final step in our calculation is to perform the sum
of $W_l^{(4+)}$ over $l$ in order to obtain the renormalized energy, and
then to study the behaviour of the renormalized energy as 
a function of $\alpha R$.  Fig.~\ref{fig:chargeen} contains
plots of both the renormalized charge and the renormalized energy
as functions of the dimensionless parameter $\alpha R$.  
This figure represents the central result of our paper.
The plots of both the charge and the energy have been normalized to
the values which one obtains assuming a condensate with chemical potential
$\mu=\alpha$, Eq.~(\ref{qcond}) and Eq.~(\ref{wcond}).

For $\alpha R < \pi$, the charge is equal to zero and the renormalized energy
is well described by keeping only the first non-vanishing term
in the perturbative expansion 
[see Eq.~(\ref{e4plustrunc})].  This limit is indicated by
the dashed line in Fig.~\ref{fig:chargeen}(b).
For $\alpha R > \pi$
the energy is no longer well-described by the leading term in
the perturbative expansion,
and instead approaches the value which one expects for a large-volume
system with chemical potential $\mu=\alpha$.

Fig.~\ref{fig:chargeen} demonstrates very convincingly that
the magnitude of the shift in energy of the vacuum does not increase
exponentially  as the potential becomes large, but
rather crosses over smoothly to the value expected for
a condensate of massless neutrinos.  The 
numerical calculations become quite
computer-intensive for large $\alpha R$ and we have not gone beyond
 $\alpha R = 100$. 
In our opinion, 
there is no  compelling reason to go to
larger values of $\alpha R$. 

\section{Conclusions}
\label{sec:conc}            

In this paper we have presented an exact non-perturbative one-loop
calculation of the neutrino vacuum energy in 
 the presence of an external electroweak potential.
In order to simplify
the calculation, we have chosen to work with a spherical
square well potential with depth $\alpha$ and radius $R$.

The formal expression for the vacuum energy is ultraviolet divergent. 
This divergence is however limited to the
leading term in the expansion of the vacuum energy in powers of 
$\alpha R$ and 
can be renormalized 
using the standard methodology. 
The higher order terms are free of ultraviolet divergences. For 
large values of $\alpha R$, however, these terms are infrared divergent.
The non-perturbative resummation of these terms has been the main goal of
our paper. We refer to the resummed expression as the
renormalized vacuum energy.

The perturbative  expansion of 
the renormalized vacuum energy is reliable for small values of $\alpha R$.
For $\alpha R>\pi$, the perturbative expansion breaks down and we find that
the ground state contains a non-zero neutrino charge.  
The neutrino charge may be calculated exactly
for any $\alpha R$ in our model. As $\alpha R$ increases, 
the charge 
is well-approximated by the charge  of a neutrino 
condensate 
with chemical potential $\alpha$. 
The onset of the charged vacuum at $\alpha R = \pi$ is accompanied by
a smooth cross-over in the vacuum energy. (See Fig.~\ref{fig:chargeen}.)
For large $\alpha R$, the energy  approaches the value expected for
a neutrino condensate.

It has been argued in Refs.~\cite{fb,fb2} that
long range forces due to the exchange of massless neutrinos
could destabilize a neutron star. Given that a finite density of
neutrons gives rise to an effective electroweak potential for
the low energy neutrinos we can directly compare our results with these claims.
According to~\cite{fb2}, a dense microscopic system of neutrons
with radius $R \sim 2\cdot 10^{-5}$ cm and $\alpha \sim 20$ eV contains
as much energy due to the exchange of massless
neutrinos as the
total  mass of a  neutron star of radius $10$ km; {\em i.e.}, 
$W \sim 10^{66}$~eV.  By way of contrast, the infrared-sensitive
terms in our calculation contribute $W \sim  - 2.3$ keV per
 massless neutrino species. 
As emphasized previously
by Abada, {\em et al}~\cite{abada}, 
the erroneous results obtained in~\cite{fb,fb2} are due to an improper
counting of the processes contributing to the vacuum energy. 
When properly resummed, these processes lead to small effects, even
for massless neutrinos,  and do not
endanger a neutron star.

\section*{Note added in proof}

In this note we 
address some issues raised in two manuscripts~\cite{fischxxx,penexxx} which
appeared after the completion of the present work.

In Ref.~\cite{fischxxx},  
Fischbach and Woodahl have questioned the validity
of the effective field theory approach used in the present paper. Their
criticism is based on a discrepancy between our
estimate of the two-body contribution to the vacuum energy of a 
neutron star [see the discussion below Eq.~(\ref{rene})], 
\begin{equation}
\label{lin}
W^{(2)} \sim \alpha^2 R^2 \Lambda,
\end{equation}
and their estimate [Eqs.~(22) and (23) in~\cite{fischxxx}],
\begin{equation}
\label{quad}
W^{(2)} \sim \alpha^2 R^3 \Lambda^2.
\end{equation}

It is however straightforward to reconcile these estimates.
The discrepancy may be traced to the fact that we have used a 
smooth potential to
model the star. 
This mean field approximation, while very well-suited for studying
 the infrared
behavior of the system, is a poor approximation to a realistic star in the
ultraviolet, where the coarse grained structure becomes important.
It is easy to verify that upon introducing fluctuations on  scales 
of order $1/\Lambda$ and repeating the analysis of Section~III.C, one
obtains an estimate for $W^{(2)}$ which is in 
qualitative agreement with Eq.~(\ref{quad}).  

We reiterate, however, that the higher-point
contributions to the vacuum energy [i.e., the terms with
$k \geq 4$ in Eq.~(\ref{exp2})] are  
insensitive to the ultraviolet features of the star. 
As discussed at length in the text, these 
contributions must be resummed, 
a task for which the effective field 
theory approach in the mean field approximation is correct and
perfectly suited. 
The resummed result, $W^{(4+)}$, tends to the 
value expected for a condensate of neutrinos with
chemical potential $\alpha$ as $\alpha R$ gets large.  This
quantity gives the leading contribution to the vacuum energy for a
smooth potential. For a
realistic star, the dominant contribution to the vacuum energy  is 
likely to be due to the two-point contribution.

In Ref.~\cite{penexxx},
Abada, P\`ene and Rodriguez-Quintero
have independently  analyzed the behavior of massless neutrinos in
a
three-dimensional electroweak potential
and have, like us, concluded that
the exchange of massless neutrinos does not threaten the 
stability of a  neutron star. 
However, their estimate of the vacuum energy,
\begin{equation}
\label{cubic}
W \sim \alpha \, R^3 \Lambda^3,
\end{equation}
differs both  from our two-body contribution~(\ref{lin}), valid for a
smooth potential,  and from our 
estimate of the contribution due to the  higher-point diagrams,
\begin{equation}
\label{condense}
W^{(4+)} \simeq  W_{\rm cond} = - {\alpha^4 R^3\over 18 \pi}.
\end{equation}
We believe that the estimate in Eq.~(\ref{cubic}) is incomplete.
Note that this expression is linear in the external potential $\alpha$. 
Changing the sign of the external potential is, however, 
equivalent to considering an antineutron star (which attracts antineutrinos)
instead of a neutron star (which attracts neutrinos). This transformation
leaves the vacuum energy unchanged. That the vacuum energy
is an even function of $\alpha$ is indeed manifest to all orders in the
Schwinger-Dyson expansion. In particular, the tadpole diagram
[which behaves superficially like  the expression 
in Eq.~(\ref{cubic})] vanishes identically when 
 both the positive and
negative energy eigenstates are taken into account in the summation
over modes. The first non-vanishing contribution is then the quadratic term as 
discussed above.

\acknowledgments
We wish to thank  M. Gyulassy, 
S. Nussinov, F. Paige, R.D. Pisarski, 
J. Sapirstein and N. Weiss for useful comments. We are particularly 
indebted to R. Jaffe for many insightful and helpful conversations.
This research was supported in part by the U.S. Department of Energy
under contract number DE-AC02-76CH00016.
K.K. is also supported in part by the Natural Sciences and
Engineering Research Council of Canada.

\end{document}